# Self-Care Practices in the Context of Older Adults Living Independently

BRIDGET CASEY, University of Queensland, Australia
GREG MARSTON, University of Queensland, Australia
DHAVAL VYAS, University of Queensland, Australia

Supporting practices around self-care is crucial for enabling older adults to continue living in their own homes and "ageing in place". While existing assistive technology and research concerning self-care practices have been centered on a medicalized viewpoint, it neglects a holistic perspective of older adults' preferences in self-care. This paper presents a study involving 12 older adults aged 65 and above in a semi-structured interview study, where we aimed to understand participants' practices around self-care. Our findings show that self-care in such cases involves activities across the physical, emotional and psychological, social, leisure and spiritual domains. This paper provides a comprehensive understanding of the daily self-care practices of older adults including an updated self-care framework identifying key aspects, and a set of design implications for self-care assistive technologies.

CCS Concepts: • Human-centered Computing • Empirical Studies in HCI

KEYWORDS: Self-care, CSCW, Older adults

## 1 INTRODUCTION

With advancements in medicine and technology, people worldwide are living longer. As a result, the number of older adults and their proportion of the population is increasing in every country across the globe. In Australia, the number of adults aged 85 years and over has increased by 110% over the past two decades, while the total population has increased by only 35%, and this trend is expected to continue [51]. Worldwide, the proportion of population over 60 years old is projected to nearly double, from 12% in 2015 to 22% in 2050 [53]. As the number of older adults grows at a significantly faster rate than the rest of the population, so too will the dependency ratio – this is a ratio of people aged outside the traditional working age range (15 to 64) compared to the working-age population [50]. An increase in the dependency ratio will be particularly apparent in the aged care sector. The Royal Commission into Aged Care Quality and Safety, established in 2018, estimated a 70% increase in staffing, or an addition of 130,000 full-time equivalent workers,

Author's addresses: B. Casey, The University of Queensland, St. Lucia, 4072 QLD Australia. bridget.casey@uq.net.au. G. Marston, The University of Queensland, St. Lucia, 4072 QLD Australia. g.marston@uq.edu.au. D. Vyas, The University of Queensland, St. Lucia, 4072 QLD Australia. d.vyas@uq.edu.au.

will be required by 2050 to ensure there is no gap in the supply and demand for aged care workers [52].

The HCI and CSCW communities have seen great amount of work on conceptualizing ageing [31,42] not as a medical problem, but a process through which older adults' wellbeing should be at the center. Researchers have growingly realized that there is a need to move away from the stereotypes older adults are associated with [12] and treat them as active members of the society who are capable to create their own future. Light et al.'s [24] agenda for the CSCW research to develop an "understanding that reveals older people's agency in the ageing process and the work they do to manage their capacity to age well" makes it clear that CSCW researchers need to have a better handle over the practices that older adults have in place to support their ageing process.

Self-care can be seen as an important aspect in this regard as it encapsulates a set of activities and tasks people perform to support themselves. Within the HCI literature, self-care is defined as "the activities that people living with a chronic condition (patients and carers) undertake to manage the condition as part of their everyday life" [27]. This is clearly a medically focused definition, which may not capture the larger issues that older adults practice in order to take care of themselves. Fields of social work and nursing have conceptualized the notion of self-care in a much broader sense, where self-care is shown in a spectrum of caring for oneself, to sense of subjective well-being, to any activity that one does to feel good about oneself [23,26,36]. Lee and Miller [23] have developed a framework to study self-care where it has been conceptualized as a combination of physical, emotional and psychological, social, leisure and spiritual aspects.

This paper's primary aim is to better understand how older adults living independently are currently practicing self-care and how assistive technology could be designed to support these practices. We studied self-care practices amongst older adults using Lee and Miller's framework [23]. We conducted a semi-structured interview study with 12 participants living in and around a retirement village, focusing on studying self-care practices of these participants across physical, emotional and psychological, social, leisure and spiritual aspects. This paper makes three main contributions to the CSCW literature: 1) It provides an empirical account of practices associated with self-care amongst older adults. This makes a strong contribution to the field of CSCW which has its roots in understanding 'work practices' – as the work older adults do to manage their own ageing process. 2) The paper learns and contributes towards a much broader understanding of self-care that goes beyond the medicalized definitions of self-care. Our findings add to Lee and Miller's framework by providing granular understanding of self-care. 3) The paper also presents a set of design implications for developing assistive technologies to support self-care.

## 2 RELATED WORK

### 2.1 Self-Care in Older Adults

Self-care is a multidimensional concept with varying definitions across literature, where no single definition is broadly accepted [13,30]. Perspectives on self-care have also been found to differ between healthcare professionals and the public [13]. An investigation conducted by Richard & Shea [30] found 139 different definitions of self-care across literature. This study also identified terms that are often used synonymously with self-care, including self-management, self-monitoring, and symptom management. However, these are separate concepts. Self-care refers to the ability to care for oneself and the performance of activities necessary to achieve, maintain, or promote optimal health. Conversely, the other terms refer to a small portion of self-care, pertaining to the management of health conditions and the medical aspects of self-care only [30].

In medical and nursing fields, self-care studies have focused on the management of health conditions [10,15,49]. For example, Zavertnik et al. [49] defined self-care as maintaining physiological stability, symptom monitoring, and treatment adherence. A set of HCI literature exploring self-care [3,8,25,27,35,37] have revealed a heavy focus on medical aspect of self-care, often investigating self-care in the context of managing a condition such as Parkinson's disease or diabetes mellitus. In the study conducted by Nunes and Fitzpatrick [27], the broadest definition was found to be self-care as the ability of individuals to manage symptoms, treatment, emotions, and lifestyle changes as part of living with a chronic condition. This still gives a limited definition, focusing on self-care in the context of managing health conditions.

Going beyond the HCI and CSCW literature, self-care is defined as a practice of activities that individuals perform independently to maintain health and wellbeing [36]. Lee & Miller [23] present a broad definition of self-care consisting of five domains: physical, psychological and emotional, social, leisure, and spiritual. These domains encompass a range of self-care practices which promote the overall health and wellbeing of oneself, and these are summarized in Table 1.

Table 1: Domains of Self-care

| Domain | Example Practices |
| --- | --- |
| Physical | Physical activity, adequate sleep, prevention of illness |
| Emotional & psychological | Stress management techniques, recognizing one's strengths |
| Social | Participating in the community, maintaining contact with important individuals |
| Leisure | Creative pursuits, sports |
| Spiritual | Meditation, prayer, reflection |

Extensive work has been done remote monitoring of changes in an individual's condition or environment to manage the risks of independent living [33]. Self-care assistive technology available to serve this purpose includes personal and emergency alarm systems, sensors for monitoring events such as high temperature, and sensor systems for monitoring and predicting care needs [6,14]. An example of such a system was developed by Austin et al. [2] and involved motion sensors, phone monitors, and contact sensors to monitor doors opening and closing. An extensive focus is also placed on physiological sensing devices for monitoring health indicators such as blood pressure and weight [11]. Another common monitoring device is wearable fall detection devices, which utilize sensors including, but not limited to, accelerometers, position tilt switches, and vibration sensors with more sophisticated devices also providing visualizations of sensor data and algorithmic assessments of fall risk [48]. Other studies include designing technology to support the self-monitoring of health indicators such as glucose levels [3], sleep, and mood levels, medication and activity tracking [41], wearable devices for monitoring fall risk [48], wearable alarm systems and systems for monitoring motion and shower usage [8].

## 2.2 Ageing-in-Place & Older Adults in HCI/CSCW

"Ageing in place" refers to older adults remaining living within the community with some level of independence, as opposed to living in residential care [47]. The concept of ageing in place has been around since the 80s and incorporate five aspects: place, social networks, support, 26:4	Bridget Casey et al.

technology and personal characteristics [28]. Supporting or improving the ability of older adults to continue to practice self-care has been an important focus of the concept ageing in place [8,19]. This concept has a strong focus on housing, ensuring the home is set up appropriately to increase safety and ensure accessibility [47].

HCI and CSCW literature has seen tremendous growth of research involving older adults. It is now well established that ageing needs be seen as a process and not a problem that needs to be fixed [42], where researchers need to get away from the existing myths and stereotypes about ageing [12] and think about empowering ageing people rather than helping them [31]. Several studies have shown that older adults are active contributors in various activities, involving themselves in DIY and making [9,16,22,43,45], creating contents [7,21] and co-developers of innovative technologies [1,5,32]. Studies have also elicited several challenges that older adults face in terms of social isolation [5,29], engaging with complex technologies and discontinuing them because of their lack of usefulness or complexity [40,46]. For older adults living independently, studies have shown that building connections and engaging in leisure activities helps them towards successful ageing [20].

Work that is put in by informal care providers has gained a lot of attention in CSCW [39]. Schorch et al. [34] have argued that care givers face significant challenges in managing their ongoing tasks with limited support and resources. Karusala et al. [18] argues for a more equitable and inclusive approach to care work within CSCW, and need for overcoming systemic inequities faced by care providers. Often, family members and close relatives play the role of informal carers. Tixier and Lewkowicz [38] show that informal carers value social support on online and offline groups where they can share experiences and learn from others and feel that they are not on their own. While our study does not involve engaging with carers directly, the above studies have informed our work on understanding larger challenges in the care ecology.

## 3 Methodology

### 3.1 Semi-structured Interviews

To gain an in-depth understanding of the everyday lives of older adults, the self-care activities they engage in, and the challenges they face related to self-care tasks, we applied in-situ, semi-structured interviews in participants' homes. The interview questions were loosely structured around the five domains derived from Lee and Miller's framework of self-care [23] and included open-ended and clarifying questions. Specific questions were posed to gain an understanding of how the individual defined and practiced self-care, as well as more general questions such as "What does a typical day look like?" to broaden the conversation and record further information. As the interviews took place in participants' homes, we were able to ask questions based on our observations and contextual cues.

Interviews started with us explaining the overall aims of the project and getting participants to sign the informed consent form (approved by our institute's ethics committee. The consent form clarified the voluntary nature of their participation in the study and that the study will only utilize de-identified data to ensure that their privacy is protected.

In total, 12 participants were interviewed resulting in around 10 hours of audio-recordings. Audio recording was chosen over notetaking, to ensure full attention was provided to the participant, as it is less intrusive. Several photos were also taken in the participants' homes to capture information related to their practice of self-care, including capturing their home setup and equipment in use.

**3.2 Participants**

The participants in the study were older adults aged between 65 to 97, with an average age of 78 years old. As the aims of the study was to investigate self-care practices in older adults, it was important to engage participants who were living by themselves. We engaged with a local retirement village that was situated in a metropolitan city in Australia. Using an 'opt in' approach, we placed our flyer in the common area of the retirement village, where participants were able to make their decision of joining our study. From this retirement village, we were able to recruit nine participants and through word of mouth we added three more participants in our study, who also lived by themselves. The details of the 12 participants are shown in Table 2 using pseudonyms. At the completion of our interview, each participant received a $20 gift voucher.

Table 2: Participant details (using pseudonyms)

| Name | Sex | Age | Dwelling Type | Known Health Conditions |
|---|---|---|---|---|
| Jack | M | 70s | Retirement village unit | Knee osteoarthritis |
| Mary | F | 70s | Retirement village unit | Previous spine fractures, wrist osteoarthritis |
| Pat | F | 80s | Retirement village unit | Previous brain tumour, knee replacements, osteoarthritis in hands, asthma, type II diabetes |
| Barb | F | 70s | Retirement village unit | Knee and right hip replacements, left hip osteoarthritis, disc protrusion, previous stroke |
| Betty | F | 80s | Retirement village unit | Previous spine fractures, osteoarthritis in spine, macular degeneration, hypertension, previous depression |
| Nancy | F | 70s | Retirement village unit | Recent heart surgery, emphysema |
| Kath | F | 80s | Retirement village unit | Vertigo |
| Joan | F | 90s | Retirement village unit | Previous hip fracture, pacemaker, leg ulcers, knee osteoarthritis, osteoporosis |
| Sue | F | 70s | Retirement village unit | Charcot-Marie-Tooth (CMT) disease, previous back injury and persisting pain, Lupus, previous heart failure |
| Ken | M | 70s | House | Back pain and degeneration, previous obesity, type II diabetes |
| John | M | 70s | House | Spinal disc herniations |
| Paul | M | 60s | Unit | Previous spine fractures, depression, mosquito-borne disease |

**3.3 Data Analysis**

Data analysis process involved manually coding the interview transcripts based on the five self-care domains [23], identifying themes within each domain, and refining these themes to produce a set of findings. The focus was on understanding the daily self-care practices and needs of older adults. Literature was also reviewed and compared to the findings to determine novel information. The findings were also considered in the context of the potential of self-care assistive technology.

**4  RESULTS**

This section outlines the way older adults perceive and engage in self-care, incorporating physical, emotional, psychological, social, leisure, and spiritual aspects to create a comprehensive framework. The participants' current technology utilization and openness to introducing new self-care assistive technology are also explored. Excerpts from interviews as well as photos from participants' homes are presented alongside the findings.

### 4.1 Defining Self-Care

Despite all participants reporting at least one medical condition, when asked to define self-care, only two participants defined self-care from a health or medical perspective. This included Paul, a man in his 60s who has a range of psychological and physical diagnoses, including a mosquito-borne disease and depression, resulting in numerous hospitalizations. Paul considered self-care as the ability to tend to both his mental and physical wellbeing, prioritizing his mental health significantly. Another participant, John, a former organic farmer and healthcare worker now in his 70s living alone in North Queensland, viewed self-care in terms of looking after his physical health.

> "If I do, on the odd occasion, go to the doctor once every year or two, it's only to get blood results and I crunch the numbers and make my own diagnosis and treatment."

John illustrates that, for some, self-care means taking responsibility for their own physical health through monitoring health indicators and responding to any changes. Instead of taking direction from clinicians, John did his own research and implemented appropriate actions and lifestyle changes to address any physical health issues.

Some participants defined self-care from a much broader and non-medicalized perspective and with a strong focus on independent living. This included defining self-care within the terms of an ability to remain living in their own home. One participant who defined self-care in this way was Sue, a lady in her 70s living in the retirement village, with a diagnosis of CMT causing progressive muscle weakness [64]. For Sue, self-care simply meant "that you can stay in your own home until it's time to go into a nursing home." Sue did not view receiving assistance within her home as undermining her independence, but as an opportunity to continue living in her own home, which was equivalent to practicing self-care.

Other participants viewed self-care as their capacity to perform tasks, gauging the level of assistance they required. This included Nancy and Barb, two ladies in their 70s living in the retirement village. Nancy, a current grocery store worker who had recently undergone a heart surgery, defined self-care as "being able to do everything that I possibly can." Barb, a retired nurse with a range of physical conditions including osteoarthritis and disc protrusions, described self-care as needing only "a minimal amount of help". For these participants, self-care was the ability to perform day-to-day tasks independently or with minimal help.

The tasks that participants associated with self-care were not medically focused, but were instead a broad range of general, menial tasks. For example, Mary, a lady in her 70s living in the retirement village who had just recovered from spinal fractures, defined self-care as "looking after everything myself," which included "keeping my place together" and doing "housework." This demonstrates that for Mary, self-care was about keeping herself and her house in order and encompassed a range of tasks including making her bed, tidying up, showering and putting on makeup in the morning. An extensive list of self-care activities was provided by Betty, a lady in her late 80s, who had just moved into the retirement village. Betty reported significant vision loss secondary to macular degeneration and defined self-care in terms of a list of daily tasks.

> "Checking my calendar every morning to see what I have to do for the day or what for tomorrow. Making sure I have proper meals with plenty of vegetables. … Drinking plenty of water. Oh, just in general looking after myself, having a shower every morning. And keeping clean. Going to the toilet. … It's just everyday things."

Betty illustrates how self-care can be performing day-to-day, menial tasks with little focus on specific medical conditions and treatment. Betty did not mention taking medication which is the focus of many studies investigating self-care [34]. Instead, Betty stated tasks that focused on making healthy choices, attending to personal hygiene, staying organized by doing things such as checking her calendar, and going to the toilet. This reflected a common theme among participants with a broad range of self-care tasks reported across all five self-care domains.

## 4.2 Physical Domain

The interviews conducted revealed that older adults actively engage in self-care within the physical domain through a varied and broad range of activities. These practices were found to fit within three sub-domains: engaging with health professionals, activity-modifying behavior, and performing activities associated with improving or maintaining general physical health. The participants' openness to introducing new assistive technology was also explored within the context of the physical self-care domain.

### 4.2.1 Regular Engagement with Health Professionals

All participants in the study were found to engage with one or more health professional(s), however, the type of healthcare professional they engaged with, the type of service (in-home vs. clinic) and the frequency of engagement varied according to the participants' needs and preferences. For example, as mentioned, John only visits his doctor occasionally to receive blood test results as he uses doctors "purely and simply for diagnosis". John, as well as Nancy, were both found to prefer natural medicine with Nancy stating, "I'm a strong believer in natural stuff and most of my medications are natural." As a result, Nancy also preferred visiting her doctor only occasionally - mostly to receive blood test results and medical certificates - and instead preferred to engage with a naturopath. Most participants, however, had a more traditional relationship with their doctor. For example, Barb visited her doctor regularly reporting, "I'm on quite a lot of medications. So, I have to see him to get scripts." These medications are shown in Figure 1. Several other participants reported regular interaction with their doctor and almost all mentioned taking some form of medication daily.

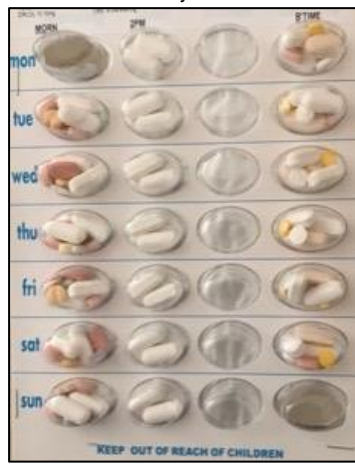

Figure 1: Barb's medication pack

Doctors, however, were not the only health professional participants engaged with, with a variety of health services accessed to support participants' practice of self-care within the physical domain. For example, Paul reported severe chronic pain and engaged with a psychologist and physiotherapist for pain relief. Paul also had an occupational therapist as he describes below.

> *"I've got an occupational therapist, which is really nice. She comes here quite often. She's the one who organized the electric bed for me and the electric lounge over there and bits and pieces. Got me set up. So, that makes life a lot easier."*

This was a common theme in interviews, with many participants reporting a positive experience engaging with health professionals. As with Paul, these professionals provided participants with services and equipment, including self-care assistive technology, which improved quality of life and supported the ability to practice self-care.

### 4.2.2 Comfort and Convenience through Activity Modifying Behavior

A common theme that emerged from our data was activity modification to reduce physical discomfort and maintain the ability to complete daily tasks. Activity-modifying behaviors include limiting involvement in activities due to loss of functional capacity, continuing activity by enhancing one's resources to avoid the aggravation of medical conditions/injuries and using new ways to achieve goal.

Throughout the interviews, participants reported a range of strategies they use to continue to practice self-care while avoiding, limiting, or modifying activities that cause aggravation of pain or other symptoms. For example, Jack, a man in his 70s living in the retirement village, has knee osteoarthritis causing significant pain when standing and, as a result, has developed a range of strategies for maintaining function and improving comfort throughout the day. He reported, "You devise ways of doing things with as little effort as you can." For example, Jack doesn't go shopping as he can't walk up and down the shopping aisles. Although he does have a mobility scooter, he reported that taking the scooter apart and putting it in the car and then having to reassemble it at the other end was "all too much". Instead, Jack has his groceries delivered, "I just get online, place my order, pay for it online and they deliver it." To put his groceries away he uses a stool, shown in Figure 2, to avoid standing for long periods.

> *"There's a little seat there in the kitchen. It's got wheels on it. I just sit on that and wheel myself around from the fridge to the sink or whatever needs to be done. And I can pack all that stuff away in the cupboards and the fridge and everything else."*

Jack engages in activity-modifying behavior to avoid standing for long periods and aggravating his knee pain. In doing so, Jack maintains the ability to shop for groceries while also avoiding pain, improving his daily comfort levels. Like Jack, several other participants reported difficulty with either grocery shopping or cooking and had developed strategies to overcome these limitations. For example, Sue reported difficulty with grocery shopping and food preparation due to foot pain related to her CMT diagnosis and orders Lite N Easy meals as a result. If Sue does have to go grocery shopping for smaller items, she goes with her daughter and uses her wheelchair, shown in Figure 3.

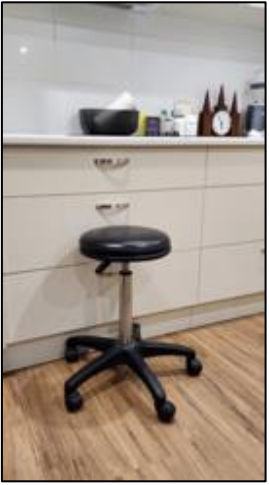 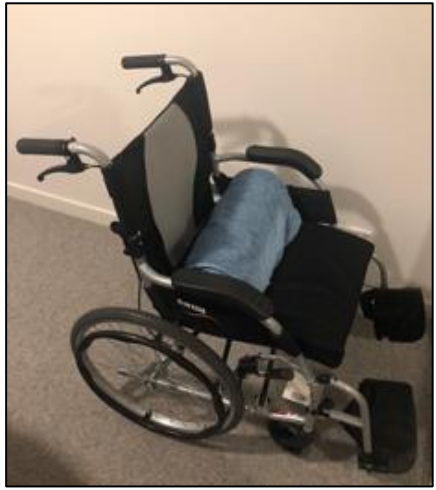 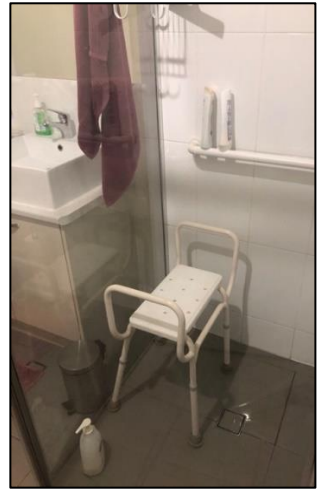

Figure 2: Jack's stool.          Figure 3: Sue's wheelchair          Figure 4: Joan's shower chair

Eight other participants reported modifications to grocery shopping and meal preparation tasks including using a mobility scooter at the shops, having a family member complete grocery shopping for them, cooking in bulk and freezing food to reduce the number of cooking sessions required, receiving Meals on Wheels, or buying ready-made meals. With these modifications in place, participants were able to continue being able to feed themselves without aggravating pain and other symptoms. The strategies differed across participants reflecting different needs and limitations. Participants such as Jack used a dryer, instead of hanging clothes outdoors after washing them. Jack understood he was unable to hang the washing out due to the standing time required but was still able to independently perform laundry by using a dryer, avoiding aggravation of knee pain, and successfully completing the task.

Alongside grocery shopping, meal preparation, and laundry, cleaning was also an activity that many participants reported difficulty with. To overcome this, the use of cleaning services were common with all but two participants receiving a weekly to fortnightly service. Participants often performed some cleaning tasks themselves but avoided activities that aggravated pain and other conditions. This included Pat, a lady in her 80s who has lived in the retirement village for 10 years.

> "I find if I've got to do the vacuuming, you know, you get a sore back and that. So, it is lovely to have the help."

This shows difficulty with specific tasks such as vacuuming due to the aggravation of pain, which is alleviated by avoiding the task all together and getting assistance. This assistance was much appreciated by Pat. Although family assistance was received by some participants, the conducted interviews revealed most help received was through paid services.

The other main area participants reported difficulty with was attending to personal hygiene, including showering and dressing, with almost all having some combination of a shower chair, stool, non-slip mat, or handrails in their bathroom. For example, Joan has a stool and rail in her shower, and uses a seat to dry and get dressed, with this equipment shown in Figure 4 and outlined by Joan below.

> "I've got a shower chair. Or stool. And I've got a plastic chair in there to sit now. I never used to. ... If I get a bit wobbly, I just grab a rail".

Here Joan had not previously used a chair in her bathroom but has adopted one recently due to her balance gradually declining abilities. If she faces balancing issues during shower, she can use the grabrails in her bathroom to steady herself. Joan also demonstrated a preference of using equipment over having a personal carer assist with these personal care tasks.

> "They [Care Agency] wanted someone to come in and shower me. I said, 'I don't want anyone to come in and shower me, I'll shower myself.' 'Oh, well what about somebody coming in and being in the house?' I said, 'look, if anything's going to happen, I'll press the button [a personal alarm system], and I'll get help.' ... And she said, 'well, what about somebody washing your hair?' I said, 'love, I'll wash my own hair. In fact, I washed it this morning.'"

Here, Joan is demonstrating resistance to an aged-care worker coming into her home to assist with tasks such as showering and washing her hair. She instead prefers to perform these tasks independently with the assistance of equipment. Joan is proud of her ability to continue to complete these tasks.

**4.2.3 Self-managed Physical Health**
Our findings revealed that nearly all participants were actively looking after their physical health through a variety of means. The most common physical health activity was exercise, with most participants consciously keeping active by regularly walking, gardening, and/or completing strength exercises.

Joan, a retirement village resident, at 97 years of age with a range of physical conditions including osteoarthritis and osteoporosis, continues to complete her daily exercises. Joan spoke about her daily routine, which involved walking up to the mailbox and around the village a few times a day. This pattern was also common among six other participants who reported regularly taking purposeful walks in their community for physical wellbeing.

Gardening was also a form of exercise reported by many participants. For example, Ken aimed to stay active by gardening daily.

> "I'm out there for hours - raking and weeding and doing well, pruning and climbing and hosing, watering."

For Ken, as well as being an enjoyable activity, gardening was performed deliberately for physical exercise as it involved a range of activities that required physical exertion as outlined in the above quote. Ken, along with many other participants, also looked after their physical health through diet. Ken placed high importance on diet for controlling both obesity and type II diabetes. He reported, "The only way I could control my obesity is to kind of just shop for the very next meal." As well as portion control, Ken also had a special diet he developed with the help of a community health center.

> "I wanted a very low carbohydrate diet 'cause I'm type II diabetic. That's what controls my diabetes, my diet."

This demonstrates a conscious decision by Ken to adapt his diet and manage his diabetes indicating he cares about his own health and takes personal responsibility for it. Diet to improve or maintain physical health was a common theme across the interviews with several participants reporting cooking and/or eating healthy meals. This included John who even went to the effort to grow his own food to avoid pesticides and eat organically.

> "My supermarket's really the garden. I'm an organic grower...so I've got a tropical permaculture organic food forest."

Both John and Ken demonstrate a desire to take personal responsibility for their health and do so through diet. John puts great effort into this, growing a range of fruit and vegetables in his own garden. A range of other physical health activities were performed regularly by participants including checking blood pressure using the device, using an Apple Watch to monitor heart rate, using an arm massager, leg elevation, using a foot circulation machine, sitting in the sun for Vitamin D, and using a Shakti/acupuncture mat. Barb did not wear her emergency pendant but instead wore an Apple Watch that had an inbuilt fall detector, calling the ambulance upon detection of a fall.

## 4.3 Psychological and Emotional Domains

From our interviews, two overarching themes emerged concerning the practice of self-care in the psychological and emotional domains. These were: 1) maintaining a positive mindset and 2) performing activities aimed at improving or maintaining mental health.

### 4.3.1 Positive Mindset

Our participants demonstrated strong, positive mindsets with regular performance of self-motivation. In general, participants held positive views of their lives, for example, Nancy reported "I'm really fortunate. Very fortunate." Nany described her life as "full," "active," and "a lot of fun". This positive view held by many participants was reflected in how they approached each day. Barb commented,

> "I like to be dressed properly and I like, when I'm going out, I like to put a bit of makeup on. And I think that's important because it's all too easy to think to yourself, well, what's the point in getting up and having a shower and putting clothes on? I can just lie in bed. And I think I took that from my father because my father was, he never wasted a day in his life."

Barb demonstrates here that she continues to put effort into her daily life, including her appearance, to feel good about herself. She sees this as continuing to live life to the fullest and draws inspiration from her father to continue to approach life this way. Despite their older age, participants continued to view life with positivity, aiming to stay engaged and active. Mary also reported not passing up on opportunities. She commented,

> "I want to be able to live my life to the fullness that I can at this stage. If someone sings out and says, you want to go to lunch today? Yep. Count me in. I'm in. ... sometimes my brother will ring me up from Maroochydore and he'll say, "You coming up?" Then if I've got nothing on yeah, I'll get in the car and I'll go."

Mary continuing to stay active and make the most of her opportunities demonstrates the power of a positive mindset. Another theme that emerged was determination. Instead of discussing limitations faced in life due to age and associated conditions, participants discussed continuing to function despite them.

One common practice that emerged was self-motivation to perform said tasks in the form of self-talk. For example, Betty motivates herself to walk every morning telling herself "if you don't use it, you'll lose it." Barb also spoke of self-motivation reporting "you need to stimulate yourself sometimes." Barb reported that, with joining in with activities, "sometimes you think I can't be bothered. But then I think, no, you get up and move!" Barb specifically spoke of a community group that she was a part of.

This further demonstrates the influence of mindset on behavior highlighting the importance of the psychological and emotional domain of self-care on other self-care domains. For example, the community group Barb attended had both social and spiritual elements, however, Barb reported determination and motivating self-talk was often required for her to attend.

**4.3.2 Mental Health Activities**

Participants discussed several activities they performed for their mental health. Two participants reported periods of feeling depressed and were able to identify the activities that helped them through these periods. Betty reported poor mental health secondary to her declining vision. She commented,

> "I've got a machine that reads a book to me now. I think I was starting to get a little bit depressed up until about four or five weeks ago. I was sort of pretending I wasn't depressed, but I think I might have been. 'Cause I was so frustrated I couldn't read a book or read anything properly. And I felt really bad about that."

Here Betty is explaining the significance of receiving a machine that plays audiobooks and the positive effect listening to these books has had on her mental health and alleviating symptoms of depression. Paul reported a history of clinical depression and performed several tasks to actively care for his mental health including watching movies, listening to music and going out on his mobility scooter. He reports regularly using his scooter for collecting cans in the community.

> "I can feel whatever's coming on when it's coming on and how. And I know better than to put myself into a dark area. And if I do, I try to get myself out of the dark area. And the only way I do that is to go and jump on my scooter and get the hell out of there. So when you're out in your scooter, you're out of that dark area because you're concentrating on everything else."

Here, Paul is describing his ability to understand when his mental health is declining, or has declined, and how he takes actions to improve his situation. One important activity for him is going out in the community on his scooter, and focusing on driving and his surroundings, to help his mood. Although no other participants spoke directly of their mental health, some participants were able to identify activities they performed when feeling stressed or worried. These included spending time with their pet, taking medications (curcumin and magnesium), having a glass of wine, calling family, and meditating. Several participants also reported taking no action when experiencing negative emotions. For example, Pat reported, "when I'm worried or stressed, I actually like to keep things to myself." And on a similar note, when asked what she does when feeling worried or stressed, Kath responded:

> "There's nothing much you can do. That's it. Because you're here and you just got to sort it out."

Here Kath is referring to living alone and feeling as though she must deal with her emotions by herself. She also voices feelings of powerlessness in regard to her ability to take action to deal with worry and stress. Barriers to older adults accessing mental health support include reduced mental health literacy and stigma associated with mental illness [54]. The previous quote from Betty reflects this as she reported, "I was sort of pretending I wasn't depressed," and that she was "frustrated" and "felt really bad" about her condition, indicating feelings of denial, shame, and/or guilt. This was also reflected by the significant number of participants who reported they did not reach out when potentially requiring mental health support.

### 4.4 Social Domain

All participants in the study were actively practicing self-care within the social domain through regular social interaction with relationships of varying natures. One significant relationship type that emerged was peer networks. Loneliness will also be discussed as, despite regular social interactions, all participants lived alone, and loneliness was commonly reported.

#### 4.4.1 Regular Social Interaction

Every participant was found to engage in some form of social interaction throughout their regular week. The most frequent social interactions occurred between neighbors followed by family members and friends. Neighbors were considered as friends by some participants, however, friendships with non-neighbors were also commonly reported. Social interactions occurred in person, both in participants' homes and in the community, as well as over the phone. It was common amongst most participants with many reporting numerous friendships and regular interactions with their friends. For example, Jack reported an extensive friendship circle.

> "I was involved in radio. ... I've still got friends there. So, I meet up with friends from the radio stations. I meet up with friends from my teaching career and my background there ... So, I've got lots of friends."

This shows Jack has maintained his friendships from past engagements, including work in radio and teaching, and continues to interact with these friends. He also spoke of the importance of these friends stating, "there's always somebody who's phoning up for a chat, which is good. Otherwise, I would never survive." This shows that, for Jack, an important part of self-care and maintaining wellbeing was regular social interaction and connections with others, including through regular phone calls. This was also reflected when Jack spoke of joining a Men's Shed.

> "I joined the men's shed next door when I moved here and that has been a bit of a boon to me because while I don't do much in terms of woodwork, metalwork, that sort of thing at the men's shed like a lot of other people do. I use it for communication and talking to people and basically keeping myself sane."

This further demonstrates the importance of regular social interaction for Jack as part of self-care. Community groups were found to be a common means through which participants interacted with others, with all but two participants belonging to a community or social group. Pat, for example, was involved in National Seniors.

> "I go to about five functions for them [National Seniors] during the [year]. I belong to their garden club and birthday clubs and going out for dinners. ...There's so many things on the list we can do if we've got the time to do it. And we do, we do like going."

Pat's quote demonstrates the regular social interactions that these community groups provide, which, for Pat, included going out for dinners as well as attending functions and garden club events. These garden club events involve going to morning tea at a range of different nurseries. Pat also expressed enjoyment in being involved in such a group and attending such events.

Friends and neighbors were the primary sources of social interaction for most participants. On top of this, some participants also reported regular interactions with family members. This included Sue who reported regular interaction with her daughter, Sophia.

> "When Sophia has some friends over or we have birthdays or something, I'm always at her place for a barbecue. Mother's Day, I went to my son's place … we had about 12 people for lunch, and it was really nice. I do get out and about."

This shows that Sue is included in the social events that her daughter holds including with Sophia's own friends. This allows Sue to regularly socialize with a range of people. Other participants also reported social interaction with family, but this more commonly involved one on one interactions where family members often came over to aid, rather than for social reasons. Most participants, however, received formal support rather than support from their families and reported much less frequent contact with their family members. For one participant, Paul, this interaction with his formal support providers was a significant part of his social life and he reported going on regular outings with them including going on shopping and fishing.

> "Well my social [life] is with my support workers … because all my support workers, most of them I get along with really, really well."

For Paul, his most regular and meaningful social interaction occurs with his support staff, and he describes these relationships positively.

**4.4.2 Peer Support Networks**

An important relationship and source of social interaction that emerged from the interviews was peer support networks. Most participants had strong connections with their peers, forming relationships that involved reciprocal actions and providing emotional and physical support. This theme was particularly evident in this study as nine of the participants resided in the same retirement village. One of these residents was Mary who spoke of a friendship group in the retirement village that she belonged to.

> "It took me a while to adjust [to living in the retirement home]. But yes, I've made a few good friends so I'm feeling more comfortable here… There are a few of us that sort of get together and go to lunch and do different things, which is good… We'd catch up at least once a week."

Here Mary is describing the friendship group and the regular interactions they have with each other. The reliability of the network is also evident with Mary confidently stating that they catch up at least once a week. As well as providing company, this group also provides reciprocal support. For example, Mary reported that after she fractured her back, the group "were so supportive … I couldn't have wished for any more help. They've been so good." Mary also reported that this group kept an eye out for each other and had developed an unspoken language, especially between Mary and her direct neighbor Pat.

> "Probably some days I don't see anyone, but most days there's somebody around, who just pops in. … we've got a thing. If my blinds up, I'm awake. If her [Pat's] curtains are drawn, she's okay. And she goes to church on Sunday. Well, her car was there on Sunday. And I'm like, right. So, there's that sort of thing in here that people care. And they check on you."

Here, Mary is further demonstrating the supportive nature of the friendship group, who regularly "pop in" to check up on each other. She also reports an unspoken language, which included drawing her curtains/blinds in the morning to indicate to Pat that she was awake and okay. Mary also reports knowing Pat's schedule, including regular weekly activities such as attending church, and noticing changes in regular behavior and ensuring she checks in if something does not seem right. Typically, the group tends to connect or touch base with each other during the evening if they haven't already communicated throughout the day. This demonstrates a supportive network in which older adults actively watch over each other and pay attention to any signs of trouble.

These peer networks held significance, as another prevalent theme arising from the interviews was the participants' perception of their families as occupied and their reluctance to cause any inconvenience. For example, Betty sees her daughter weekly but does not see her other children or grandchildren due to their busy schedules.

> "I see my daughter that takes me shopping every week. But if I need anything extra, she'll come. But see, she's got all these grandchildren herself. And she's busy looking after her husband and the business that they have. They're all working. All got jobs. ... I don't see them [grandchildren] very often because they're all working. They've all got children of their own and I can understand that."

Here, Betty is explaining that she sees her daughter weekly to go grocery shopping but does not expect any more than this due to how busy her daughter is with work and her own family. She also notes the infrequency of meetings with other family members, such as her grandchildren, due to their numerous commitments. The quote suggests Betty feels that visiting her requires time and effort that her family do not have the capacity to give, and she does not feel as though she should expect it.

### 4.5 Leisure Domain

Our participants were observed to regularly participate in leisure activities, primarily in solitary and home-based settings. These activities broadly fell into two categories: hobbies and activities related to rest and relaxation. A hobby refers to a set of activities participants had a passion for, dedicating time and effort, and placed a high value on. Rest and relaxation activities on the other hand referred to activities participants performed mainly to avoid boredom with relaxation and passing time.

#### 4.5.1 Hobbies

Several participants in the study engaged in hobbies. For example, Jack, born and raised in Ireland, had recently regained his love for music. He commented,

> "I started playing when I was about 10. And I've been playing bits and pieces. Lots of long times that I don't play. ... And then I came back to it again in my retirement and I do it now. My main interest is Irish music, Irish folk songs, ballads, that sort of stuff. I like those."

As well as playing music, Jack also had a mini recording studio in his home, shown in Figure 5, where he records and produces music, "I record my own music as well. I record my own playing, my own singing. I'll put them together."

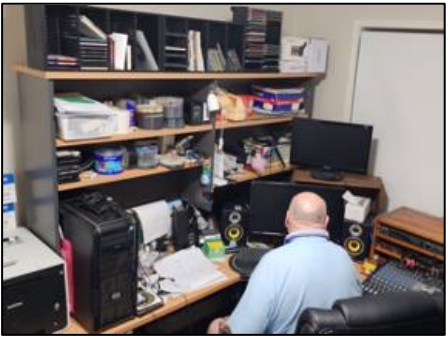 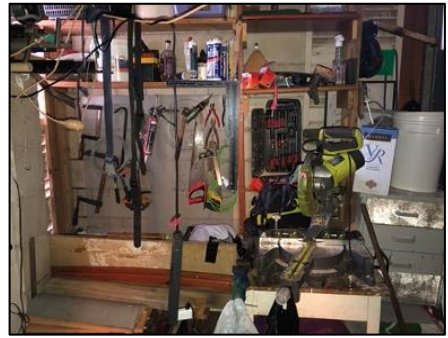

Figure 5: Jack's in-home recording studio                       Figure 6: Paul's workbench

As demonstrated by Jack's comments, music was an activity he was passionate about and linked to his Irish heritage. Not only does he play music, but he has also purchased a wide range of equipment and set up a small studio, displaying dedication. Many other participants spoke of some of their leisure activities passionately, which included gardening, embroidery, tinkering/building, family history, and flower arranging. For example, Paul enjoys building and fixing broken furniture and electronics in his home. He commented,

> "I like making things and I just like finding things to make. But I'm finding it harder and harder every time to do that. So, you've got to have some sort of drive, initiative to do it."

This shows that Paul does not just make and fix things as it is convenient or to pass time but because he truly enjoys it and has a drive for tinkering and building. Paul speaks of this hobby enthusiastically, explaining that everything in his unit "has been reclaimed, refixed, remodeled, re-something. All these fans, I've got them off the road and fixed them all up. They all work." He also exhibited pride in his building and fixing accomplishments. He had fixed things for many people and had been gifted several items in return. Figure 6 displays Paul's workbench with a large range of tools and equipment.

Some participants mentioned their inability to continue participating in previously enjoyed leisure activities due to underlying health or medical conditions. For example, Kath has had to stop attending ten pin bowling due to vertigo. She commented,

> "10 pin bowling, I did that for years. And I've still got my bowling ball there. I'd love to do it. But with the noise and with the ear and that, it's just, it's out of the question."

Here, Kath is explaining that she can no longer bowl due to the loud noise at bowling alleys. This is because vertigo and imbalance can be triggered by loud noises in the affected ear [71]. Kath still wished she could perform this activity and is holding onto hope, keeping her bowling ball in her possession. The above quote was in response to Kath being asked if there were any activities she would perform if worried or stressed. This further highlights the importance of this activity for Kath. When asked if there were any activities that Kath thought of when she thinks of self-care, she also thought of 10-pin bowling.

> "No, not really, because as I said, since I've got the vertigo, I can't go bowling, which I did love doing."

Here, Kath is saying that she does not consider doing any activities for self-care as 10-pin bowling was the main activity she engaged in to care for herself, and she can no longer perform this activity. This again highlights how important this activity was to Kath. It also highlights the important role leisure and hobbies can play in a person's self-care plan.

**4.5.2 Rest and Relaxation Activities**
Although not all participants had a hobby, every participant engaged in some form of activity for rest and relaxation. For example, as well as playing and recording music, Jack also completed crosswords in his spare time.

> "I spend my time doing crosswords. I've done all the crosswords in today's paper already.
> … So that takes a bit of time. But I enjoy that."

Jack mentions his enjoyment of completing crosswords but perceives this leisure activity as more of a way to pass the time. Like Jack, many participants spoke of rest and relaxation activities they engage in to pass the time and others reported engaging in these activities to avoid boredom. For example, Pat said, "I have got games on my iPad that I, if I was bored, I could get up a game." This was a common activity and Kath's iPad showing her favorite game is displayed in Figure 7.

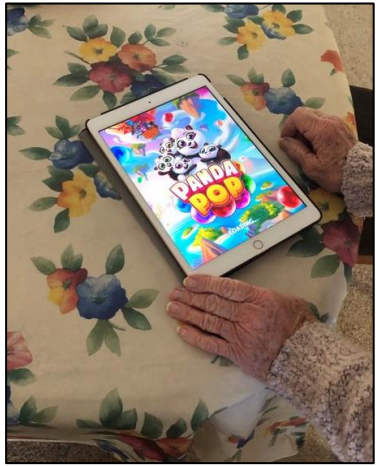

Figure 7: Kath's iPad displaying her favorite game.

The activities participants engaged in varied across individuals with puzzles and games being common as well as watching tv and movies, reading, cooking, and drawing. Pat had hobbies which included gardening and embroidery and also watched tv to pass the time.

> "I've got Netflix and a few things that I can usually find a movie … so that takes a lot of watching… If you think, oh, what can I do? Oh, I can watch another few episodes of that."

Here, Pat is explaining that if she ever finds herself bored or wondering what to do, she can put on a movie or a few episodes of TV to pass the time. She does so using the streaming service Netflix. TV was a common activity of participants when bored.

## 4.6 Spiritual Domain

All participants were found to practice self-care within the spiritual domain by either engaging in spiritual or religious practices, engaging in altruistic activities, or a combination of both. In Richards et al.'s work on self-care [35], spirituality can be generally described as a sense of the purpose and meaning of life.

### 4.6.1 Spiritual and Religious Practices

Almost all participants interviewed classified themselves as either spiritual, religious or both. Nancy described her faith as the following,

> "I'm a Christian. I'm not a very active, practicing Christian, but I have my Bible and I have [religious] texts sent to me every day. … I have my faith and my beliefs."

What varied significantly, however, was how participants expressed and practiced their beliefs. For example, Mary viewed religion as the way she interacts with people.

> "I was very religious, when I moved to the country, it was just not convenient to go [to church]. I still have all my beliefs. I just don't believe I need to go to a church to do it. To me, religion is how you treat people and how you deal with people. If you do the right thing by people, you're doing the right thing."

As Mary rarely attends church, her regular religious practices include living out her beliefs in terms of her interactions with others as well as prayer. Mary reported, "I don't go to bed without saying a prayer or to sleep. I've always done that." Several other participants also practiced their religion through prayer, however, their reasons for doing so were varied. Reported reasons included asking for help, expressing gratitude, or connecting with loved ones who had passed. Others spoke of praying as required, which was also the reason one participant engaged in meditation. Unlike Mary, some participants regularly attended church, either in person or through televised services. Religious and spiritual beliefs and practices were found to be a significant component of self-care bringing comfort and grounding.

> "And if you really, really do believe, it does help you in life."

Here, Mary spoke of the significance of religion and having faith in her life. She finds it extremely useful in terms of providing guidance and, in turn, comfort.

### 4.6.2 Altruistic Activities

Several participants engaged in altruistic activities that helped establish a sense of purpose. For example, Jack participates in the training of new care staff to provide them with a client's perspective, Nancy engages in paid and charity work, and Paul gives out essentials, such as toothbrushes, to those in his community that need it and runs a communal library outside his unit (Figure 8). Participants spoke of these activities with pride, including Paul when he described the usage of his communal library/book swap. Paul has placed this outside his ground-floor unit door, which is situated next to the lifts, resulting in several fellow unit block residents passing each day.

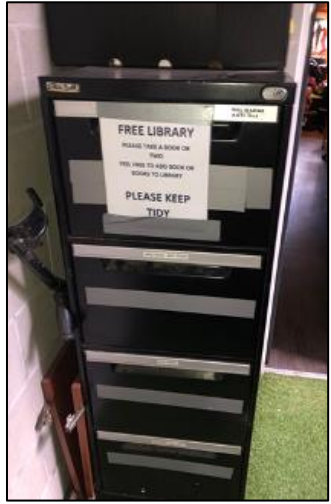

Figure 8: Paul's communal library

"[The communal library] gets used pretty well. For the amount of years it's been there, I wouldn't think that it would turn over so much, but it does. So I come out, I can hear them there of a night time, I come out and everything's different, changed around. New books."

In this context, Jack expresses his pride in the extensive use of the library. He sees it as not only contributing to the community but also fostering a cohesive and positive environment within his unit complex.

## 5 DISCUSSION

The in-depth interviews conducted in the homes of 12 older adults provided rich insights into their everyday practice of self-care. This section will discuss our findings in the context of existing literature on self-care and present design implications that emerged from the findings.

### 5.1 Understanding Self-Care

Our findings demonstrate the complex and personal nature of self-care, with participants engaging in a broad range of self-care practices, which varied across individuals. The findings also included insights into how participants themselves defined self-care. Despite their various health conditions and issues, participants' definitions of self-care had little focus on managing these health conditions and instead focused on independent living in terms of being able to remain in their own homes and their ability to complete day-to-day tasks with minimal assistance. This contrasted with research conducted in the health and HCI domains, where self-care was primarily understood as managing medical conditions [3,8,25,27,35,37]. Nunes et al.'s study [27] surveyed HCI research on self-care and found that even the broadest definition was centered around the management of health – the ability of individuals to manage symptoms, treatment, emotions, and lifestyle changes as part of living with a chronic condition.

To Richard & Shea [30], there is no consistent definition of self-care, with varying definitions used across literature. The idea of self-care between health professionals and general public and even within healthcare professionals differ vastly [13]. In the context of HCI and CSCW literature [3,8,41,48], self-care is also seen in a limited way – mainly utilizing the physical domain from Lee and Miller's framework [23]. This framework helped us view self-care in a much broader way, which include aspects of self-care that older adults value highly. As we discussed in our findings, Barb, for example, considered wearing make-up and looking good as a part of her self-care practice, which would not have been considered as a part of self-care regime within the existing HCI literature.

Our participants' definitions of self-care were more aligned with those presented in social science studies, which had a broader focus and included independent practice of any activity to maintain health and wellbeing [36]. The activities considered self-care by our participants included ten-pin bowling, housework, putting on make-up, showering, checking the calendar, drinking sufficient water, and eating vegetables. This study therefore offers a new perspective of self-care to the HCI and CSCW literature that is unique to older adults, which revealed a strong focus on independent living and the ability to maintain self-care while residing in one's home.

In Table 3, we present an updated version of the framework that Lee and Miller [23] developed. While the original framework includes physical activities, adequate sleep and prevention of illness in the *physical domain* (Table 1), our findings show that participants in our study incorporated several activity modifying behaviors to avoid body strain and avail comfort and convenience. This included some obvious practices around using shower chairs and stools to avoid falls, for example, but this also included some creative approaches of using stools with wheels when working in kitchens. Jack avoided standing up for a longer period to complete tasks in the kitchen due to inability to stand for prolonged periods. Other examples here included doing online shopping to avoid travels, using dryers to avoid lifting clothes and hanging them outdoors, and hiring help for cleaning. On the *emotional and psychological domain*, managing stress and recognizing one's strength were quite prevalent in our participants. More importantly, keeping a positive mindset came out as an important sub-domain. As shown in Barb's example, dressing up and putting makeup on, she made herself feel good. Similarly, Mary and Paul's examples show that even with their declining abilities, their 'can-do' attitude helped them live positively. On the *social domain*, one of the differentiating and notable sub-domains that emerged was peer support networks. This was particularly unique in our study since most of our participants resided in the same retirement village. Our participants checked on their neighbors based on the subtle cues of curtains being up or down and organized and participated in regular events and common activities due to their closer proximity. These connections were crucial, as numerous participants perceived their families as too occupied to provide them with social interaction or support. This aligned with previously conducted research that revealed older adults can express concerns about being a burden and can be reluctant to complicate the busy lives of their family members [55]. These relationships offer support by maintaining mutual observation and occasional check ins, along with regular social interactions, such as dining out, making visits to each other's residences, or brief exchanges when crossing paths.  This is particularly important in this age group as older adults are more likely to live alone. On the *leisure domain*, our findings were much more aligned with Lee and Miller's framework in that our participants reported of having hobbies and creative pursuits to keep themselves engaged in physical and mentally stimulating activities. However, on the *spiritual domain*, our participants went beyond the religious and spiritual practices and exhibited altruistic motivations to have a sense of purpose in their lives. For example, Paul developed a communal library to help others living this building. Several other participants such

Self-Care Practices in the Context of Older Adults Living Independently          26:21

as Nancy were involved in helping others in need. Altruistic motivations in the later life are well discussed in the CSCW literature [43] and our findings are aligned with that.

Table 3: Subdomains for self-care in older adults

| Domain | Subdomain | Explanation | Examples |
|---|---|---|---|
| **Physical** | Engaging with health professionals | Engaging in formal healthcare services. | Attending doctor's appointments, in-clinic exercise classes |
| | Activity-modifying behavior | Limiting engagement in certain activities or using new means or resources to continue certain activities without aggravating medical conditions and/or causing discomfort. | Using a shower stool for showering, hiring a cleaning service, using dryers over hanging clothes outside |
| | Physical health activities | Engaging in activities aimed at maintaining or improving general physical health. | Strength exercises, gardening, leg elevation |
| **Psychological & Emotional** | Positive mindset | Maintaining a positive mindset through regular performance of self-motivation and the practice of positivity. | Positive or motivating self-talk, putting effort into appearance |
| | Mental health activities | Engaging in activities aimed at maintaining or improving mental health. | Listening to music, spending time with a pet |
| **Social** | Regular social interaction | Regular social exchanges with other individuals. | Calling friends and family, involvement in a community group |
| | Peer support networks | Forming and maintaining networks with other older adults involving reciprocal actions and providing emotional and physical support. | Checking in on neighbors, organizing and attending regular events with friends/neighbors |
| **Leisure** | Hobbies | Engaging in activities an individual is passionate about. | Music, tinkering/building, flower arranging |
| | Rest & relaxation activities | Engaging in pleasurable activities to pass the time or ease boredom. | Puzzles, games, watching TV |
| **Spiritual** | Spiritual & religious practices | Engaging in faith-based or spiritual practices. | Prayer, meditation, attending mass |
| | Altruistic activities | Engaging in activities that established a sense of purpose. | Charity work, acts of service |

## 5.2 Implications for Design

### 5.2.1 The Potential for Self-Care Assistive Technology for Older Adults

CSCW researchers should be careful in considering whether technology would be a solution to support self-care practices of older adults. In this study, a majority of participants were open to using technologies in their homes, if they would make lives easier, fix a problem, or help participants complete tasks. This supports well-established norms that older adults are not averse to technology being integrated into their lives especially if the device is perceived as useful [4,6,32,44]. Goher et al. [14] found that older adults can be concerned about learning new technology skills, however, engaging directly with technology fosters more positive views towards its use. The study also emphasized the importance of technology interfaces being easy to use and matching the user's abilities. Similar findings emerged from Wu and Munteanu's investigation [48] which found ease of use was correlated with an overall positive user experience and high user acceptance, and Bhachu et al. [6] emphasized the importance of time and support for teaching older adults how to use devices. All these indicate that there is a potential for the successful uptake of self-care assistive technology if older adults perceive them as useful.

There has been an increasing efforts for designing assistive self-care devices that focuses on the management of health conditions [3,6,11,35,37], and therefore provide usefulness in terms of managing specific medical conditions. In doing so, self-care is considered from the perspective of, and prioritizes the needs of, healthcare professionals, rather than the older adults themselves. In our study, it was revealed that older adults viewed self-care differently, compared to healthcare professionals, and has a predominantly non-medicalized perspective with a strong focus on independent living. The activities they valued and considered to be a part of their self-care regime were also non-medicalized. This means that designing technology with a reduced focus on the medical risk and impairments of users and a stronger focus on supporting older adults in completing tasks across all five self-care domains, targeting practices they value, and promoting independence and autonomy.

**5.2.2 Self-Care Technology for Autonomy and Dignity**
Another finding from the literature review was that older adults may not want to accept that they are at risk and in need of support or care [14]. They therefore may reject technology that make them vulnerable and treats them as a 'patient'. An example of such a device is the personal alarm system, which is worn by older adults to alert ambulance services and family members if they suffer a fall. The older adult can also press the button if requiring assistance. Despite all retirement villages being fitted with an alarm system, many residents in our study were found not to wear their device as they did not perceive themselves to be at significant risk. This device is less centered on enabling older adults to maintain independence and perform tasks; instead, it primarily monitors the older adult [17]. This may clarify the low compliance witnessed and further emphasizes that self-care assistive technology needs to be designed with the older adult's autonomy and dignity in mind.

Our study also revealed that participants preferred minimal human assistance while performing self-care tasks. They expressed their preference for technology or equipment over human assistance. For example, Joan reported a preference for using shower stools and chairs over having an aged-care worker assist her with showering. Mary also reported a preference for technology over someone coming into her home to perform a service. She viewed technology as enabling her to continue to perform the task herself, suggesting technology is not seen as reducing independence or one's sense of ability, compared to receiving human assistance. Our findings are in line with Goher et al. [14], which showed that older adults prefer assistance from robots over humans for activities of daily living. We suggest that self-care technologies should be designed to support older adults according to their preferences, ensuring a sense of independence is maintained.

**5.2.3 Integrate Self-Care in Mainstream Technologies**
This study also provided unique insights into the current technology utilization of older adults. We found that older adults were already employing a range of technology to support self-care practices. For example, Barb used an Apple Watch as a fall detector and to monitor her heart rate. The acceptance of this piece of technology over the personal alarm device suggests older adults are not opposed to devices with functionality aimed at increasing safety or physical health but opposed to what the use of some devices represents. In this case, older adults saw wearing the personal alarm as a sign they were fragile or unwell. Other commonly adopted mainstream technologies included iPads for leisure, televisions for entertainment and spiritual activities, with some participants using them to watch televised religious services. Additionally, mobile phones were utilized for social purposes. To improve technology uptake, designers should consider integration with mainstream technologies already commonly used by older adults or making general technology more accessible. In doing so, older adults would not feel as if they were being treated differently due to their age and/or medical conditions, but as part of the general population.

### 5.2.4 Harness Current Positive Self-Care Practices in Technology

This study revealed several positive self-care practices performed by older adults. This included forming and maintaining peer-support networks and maintaining positive mindsets. Technology could be designed to harness these positive practices. For example, personal alarm systems tend to be set up to alert family members or emergency services, but it may be the preference of older adults to notify a member of their peer support network. Compared to relationships with family members where older adults expressed feelings of being a burden, peer-support networks were spoken of more positively, and support in these relationships was seen as reciprocal. They may therefore feel less monitored or as though they are being treated as vulnerable, using devices based on these relationships. If choosing to design technology to help connect older adults in their homes, these peer relationships could form the foundation. Older adults were also found to have strong practices around maintaining a positive mindset. Technology could be utilized to further facilitate these positive practices.

## 6 LIMITATIONS

It is important to point out that our study clearly has limitations in terms of its sample size and generalizability. Given that this study involved working with a vulnerable group of people, it would be challenging to spend more than an hour of time at the participants' homes. We understood that considering the age of our participants, spending more than an hour would create physical and mental burden on our participants. We also chose semi-structured interview method to ensure that participants are able to conduct their ongoing activities if they needed to and at the same time the in-situ aspect of our approach would lead to learning contextual aspects of our participants' everyday life. However, as we had 12 participants in our study, we believe that our findings may provide a limited understanding of self-care practices.

On the issue of generalizability, we posit that our findings present more of case-study type research than something that can be generalized. Our findings are based on specific locality from a metropolitan city in Australia. The type of government support that pensioners receive in Australia is different compared to many Western counties, hence, generalizing these findings would be far misplaced.

## 7 CONCLUSIONS

This paper has provided empirical insights around how older adults living independently are currently practicing self-care. Extending Lee and Miller's framework of self-care [23], the paper provides rich insights into the everyday self-care practices of older adults, going beyond the medicalized notions of self-care. The main takeaway from this study is that self-care assistive technology should be designed with the needs and preferences of the older adult in mind, with a strong focus on enabling independence and supporting all aspects of self-care. From the presented implications, assistive technology could be developed to assist older adults in the practice of self-care, enabling "ageing in place" as per the preferences of older adults, despite the ageing population and predicted aged-care workforce shortages. This project could serve as a valuable resource for researchers exploring the development of self-care assistive technology for various other user groups.

## ACKNOWLEDGEMENT

This work is funded by the University of Queensland's Cyber Seed Funding scheme 2022. We thank all our participants for their help and support in the study and the anonymous reviewers for their guidance.